 \documentclass[final,5p,times,twocolumn]{elsarticle}
\usepackage{lineno}

\usepackage{amssymb}
\usepackage{amsmath}
\usepackage{amsthm}

\usepackage{graphicx}
\usepackage{siunitx}
\usepackage{physics}
\usepackage{hyperref}
\usepackage{comment}
\newcommand{\um}{\ensuremath{\text{\textmu m}}}

\journal{Nuclear Physics B}

\begin{document}

\begin{frontmatter}



\title{Development of Pixelated Capacitive-Coupled LGAD (ACLGADpix) Detectors} 
\author[kek]{Koji Nakamura}
\author[tsukuba]{Yua Murayama}
\author[tsukuba]{Issei Horikoshi}
\author[tsukuba]{Mahiro Kobayashi}
\author[tsukuba]{Koji Sato}

\affiliation[kek]{organization={High Energy Accelerator Research Organization (KEK)},
        addressline={Oho 1-1},
        city={Tsukuba},
        postcode={305-0801},
        state={Ibaraki},
        country={Japan}
}
\affiliation[tsukuba]{organization={University of Tsukuba},
        addressline={Tennodai 1-1-1},
        city={Tsukuba},
        postcode={305-8571},
        state={Ibaraki},
        country={Japan}
}


\begin{abstract}
The Low-Gain Avalanche Diode (LGAD) is a semiconductor detector capable of achieving excellent timing resolution (~20 ps) for minimum ionizing particles (MIPs). To realize a pixelated detector with both high timing precision and spatial resolution, we have been developing Capacitive-Coupled LGADs (ACLGADs) for future collider experiments, such as the latter phase of the High-Luminosity LHC. We have successfully fabricated a pixelated ACLGAD (ACLGADpix) with a 100~\um{} $\times$ 100~\um{} pixel pitch, maintaining uniform timing performance across the active area.
In this presentation, we will report recent measurement results from ACLGADpix prototypes using beta rays, an infrared laser, and a 3 GeV electron beam. We will also discuss potential readout electronics for future collider applications.
\end{abstract}




\end{frontmatter}



\section{Introduction}

Future collider experiments such as the HL-LHC and proposed next-generation colliders will face unprecedented track densities due to severe pileup conditions.
At the HL-LHC, 140--200 interactions per bunch crossing are expected, while future hadron colliders may exceed 1000 interactions per bunch crossing.
Under such conditions, conventional tracking detectors relying solely on spatial information suffer from ambiguities in track reconstruction.

The addition of precise timing information at the level of tens of picoseconds enables so-called 4D tracking, significantly improving pattern recognition by separating tracks in both space and time. In future collider experiments, precise spatial information alone will not be sufficient under severe pileup conditions; timing information must be combined with high-granularity position measurements in order to associate tracks with their correct interaction vertices.

While excellent spatial resolution is already achievable with conventional silicon pixel detectors, realizing a detector that simultaneously provides fine spatial granularity and $\mathcal{O}(10)$~ps timing resolution remains a major challenge. Low Gain Avalanche Diodes (LGADs) have demonstrated outstanding timing performance, achieving resolutions of around 30~ps for minimum ionizing particles in pad-type sensors~\cite{PELLEGRINI201412,KITA_VERTEX2022,Horikoshi:2025QS,Murayama:2025oP,Imamura:2024zX,Nakamura_VERTEX2020}. However, in conventional segmented LGADs, fine segmentation requires junction termination extensions and isolation structures, which increase the inactive area. As the pitch becomes smaller, this dead area rapidly becomes dominant, making such devices unsuitable for true fine-pitch pixel tracking applications.

The AC-coupled LGAD (AC-LGAD) concept was developed to overcome this limitation by employing a continuous gain layer combined with capacitive coupling to segmented readout electrodes~\cite{KITA2023168009}. This architecture preserves full fill factor while enabling position reconstruction through segmented readout and charge sharing between neighboring electrodes. Based on this concept, we have developed pixelated AC-LGAD detectors with a pitch of 100~\um{} $\times$ 100~\um{}. In this paper, we present the first comprehensive experimental study of pixelated AC-LGAD detectors, which represents a key step from the sensor concept to a realistic 4D pixel detector.

\section{Fabrication of Pixelated AC-LGAD Detectors}

\begin{figure}[htbp]
  \centering
  \includegraphics[width=0.7\linewidth]{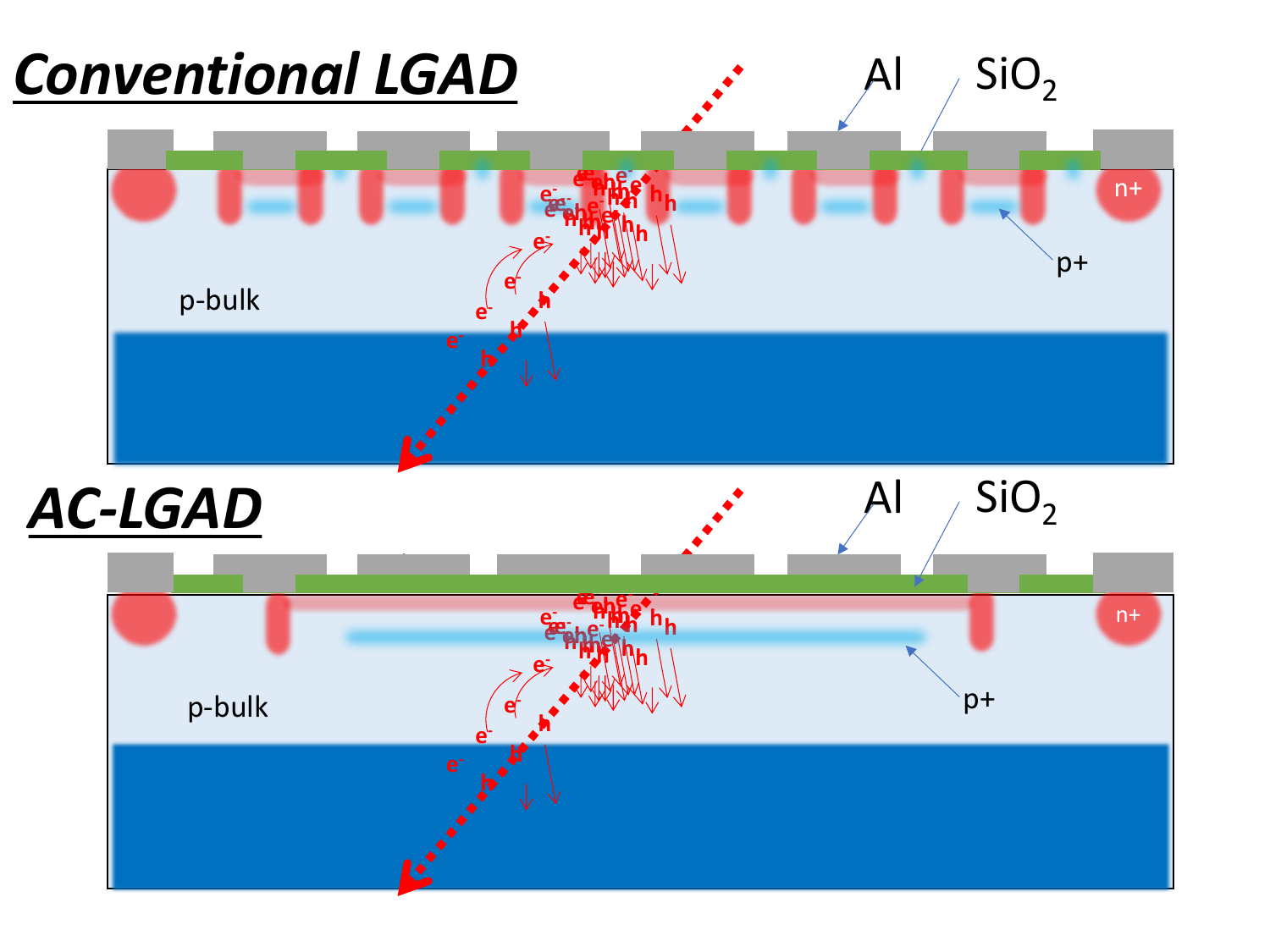}
  \caption{Comparison of the structures of a conventional segmented LGAD (top) and an AC-LGAD (bottom). In the AC-LGAD, the gain layer is continuous over the full active area and the signal is read out through AC-coupled segmented electrodes.}
  \label{fig:aclgad_structure}
\end{figure}

The pixelated AC-LGAD sensors studied in this work were fabricated by Hamamatsu Photonics K.K. in collaboration with the KEK--Tsukuba group.
Figure~\ref{fig:aclgad_structure} compares the structures of a conventional segmented LGAD (top) and an AC-LGAD (bottom).

The development of AC-LGAD technology was motivated by a limitation of conventional segmented LGADs. As illustrated in the top part of Fig.~\ref{fig:aclgad_structure}, standard LGAD structures require junction termination extensions and isolation structures between neighboring electrodes. As the segmentation becomes finer, the relative fraction of inactive area increases, making such devices unsuitable for fine-pitch pixel detectors despite their excellent timing performance.

The AC-LGAD concept was developed to overcome this limitation. As shown in the bottom part of Fig.~\ref{fig:aclgad_structure}, the gain layer is formed continuously across the sensor area, while the signal is read out through segmented metal electrodes placed above a dielectric layer. This architecture preserves full fill factor and enables position-sensitive readout without introducing dead regions associated with segmented gain structures.

The amplified signal is transferred capacitively to the metal electrodes, while part of the signal can spread laterally through the resistive n$^+$ layer to neighboring channels. The detector response is therefore governed by the balance between the coupling capacitance and the sheet resistance of the n$^+$ layer, which determines the signal amplitude, charge sharing, and crosstalk.

Based on this principle, several prototype productions were carried out to optimize the key design parameters, including the oxide thickness, the n$^+$ doping concentration, and the gain-layer doping. These parameters were tuned to obtain sufficiently large signal amplitude, stable gain at practical bias voltages, and controlled charge sharing with limited crosstalk. The resulting design, established in our previous study~\cite{KITA2023168009}, provided the basis for the pixelated sensors investigated here.

In this work, we focus on AC-LGAD sensors with a pixel pitch of 100~\um{} $\times$ 100~\um{} and an active thickness of 20~\um{}. The use of a thin active layer helps to improve the intrinsic timing performance, since the fluctuation in signal formation time caused by non-uniform charge deposition through the sensor thickness is reduced. The devices used in this work were selected from that sensor production primarily on the basis of sensor thickness and signal-to-noise ratio.

\subsection{Beta-ray Measurement Setup}

\begin{figure}[htbp]
  \centering
  \includegraphics[width=1.0\linewidth]{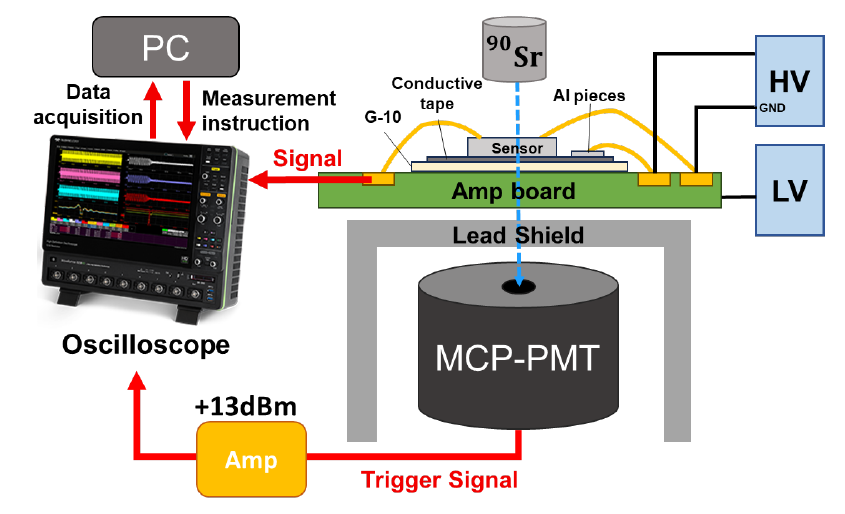}
  \caption{Experimental setup for the $^{90}$Sr beta-ray measurement. An MCP-PMT is used as the timing reference.}
  \label{fig:beta_setup}
\end{figure}

The intrinsic timing performance of the pixelated AC-LGAD detectors was evaluated using a $^{90}$Sr beta-ray source.
A schematic of the experimental setup is shown in Fig.~\ref{fig:beta_setup}.

A Hamamatsu R3809U-52 microchannel-plate photomultiplier tube (MCP-PMT) was used as the external timing reference. This device has a typical transit-time spread of 25~ps and a typical rise time of 160~ps. In our setup, its timing contribution was sufficiently small compared with that of the detector under test and was therefore neglected in the evaluation of the AC-LGAD timing resolution. 

The AC-LGAD signals were amplified with a 16-channel discrete amplifier board based on 2 step RF amplifiers. The amplified detector signals and the MCP-PMT reference signal were digitized with a Teledyne LeCroy WaveRunner 8208HD oscilloscope, which provides 2~GHz analog bandwidth, 8 input channels, 12-bit ADC resolution, and a sampling rate of 10~GS/s. 

The hit time was extracted with a 50\% constant-fraction method for both the AC-LGAD and MCP-PMT waveforms, and the timing resolution was obtained from the distribution of the time difference between the two signals. Since the MCP-PMT timing resolution is approximately 10~pico second, no correction for the reference contribution was applied.

\subsection{Electron Beam Test at the KEK PF-AR Test Beam Line}

Beam tests were carried out at the KEK PF-AR test beam line, a facility at the KEK Tsukuba campus that provides electron beams in the energy range from about 0.5 to 6.5~GeV for detector-development studies. In the present measurements, a 3~GeV electron beam was typically used ~\cite{ARTBL}.

The experimental setup included a tracking telescope composed of FE-I4 and MALTA2 pixel detectors. The FE-I4 sensors were based on the readout chip originally developed for the ATLAS Insertable B-Layer (IBL), with pixel sizes of 50~\um{} $\times$ 250~\um{}~\cite{GARCIASCIVERES2011S155}.  The MALTA2 sensors are depleted monolithic active pixel sensors with a pixel pitch of 36.4~\um{} $\times$ 36.4~\um{}, providing a much finer granularity for precise track reconstruction~\cite{SNOEYS201941}. 

An MCP-PMT was again used as the timing reference. Seven neighboring channels of the AC-LGAD pixel sensor were read out simultaneously, allowing measurements of hit position, detection efficiency, timing resolution, and crosstalk. Because the beam energy was relatively low, multiple-scattering effects were not negligible and must be taken into account when interpreting the spatial and timing performance.

The FE-I4 planes provided robust tracking over a relatively wide area, while the MALTA2 planes improved the pointing precision thanks to their finer pixel pitch. Seven neighboring channels of the AC-LGAD pixel sensor were read out simultaneously, allowing measurements of hit position, detection efficiency, timing resolution, and crosstalk.

\section{Results}
\subsection{Timing Resolution}

\begin{figure}[htbp]
  \centering
  \includegraphics[width=0.9\linewidth]{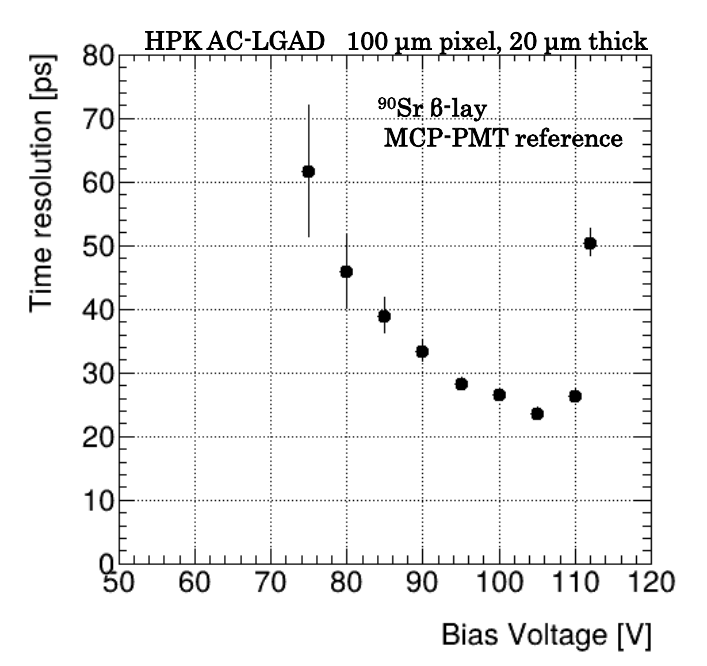}
  \caption{Timing resolution measured with a $^{90}$Sr beta-ray source for a 20~\um-thick, 100~\um-pitch AC-LGAD pixel sensor.}
  \label{fig:beta_timing}
\end{figure}

Figure~\ref{fig:beta_timing} shows the timing resolution measured with the $^{90}$Sr beta-ray setup as a function of bias voltage for the 20~\um{}-thick pixelated AC-LGAD sensor.

The timing resolution improves with increasing bias voltage and reaches its best value of
\begin{equation}
\sigma_t = 25.3 \pm 0.1~\text{ps}
\end{equation}
at 105~V. This result demonstrates that the pixelated AC-LGAD geometry preserves the excellent timing capability of LGAD-based sensors even at a fine pitch of 100~\um{} $\times$ 100~\um{}. In particular, the achieved value is comparable to that of high-performance pad-type LGADs, indicating that the pixelation does not introduce a significant penalty in timing performance.

In the electron-beam measurement, the timing resolution was found to be $40$--$45$~ps, which is worse than that obtained in the beta-ray setup. This difference is not attributed to the sensor itself, but mainly to the experimental conditions of the beam test. In particular, the relatively low beam energy at the KEK PF-AR test beam line leads to non-negligible multiple scattering, which degrades the track extrapolation accuracy and broadens the measured timing distribution. In addition, the contribution from the external timing reference and the overall beam-test system is larger than in the beta-ray setup. Therefore, the beta-ray result provides the more direct estimate of the intrinsic sensor timing performance, while the beam-test result should be interpreted as the performance of the full measurement system under realistic test-beam conditions.

\subsection{Detection Efficiency}

\begin{figure}[htbp]
  \centering
  \includegraphics[width=0.9\linewidth]{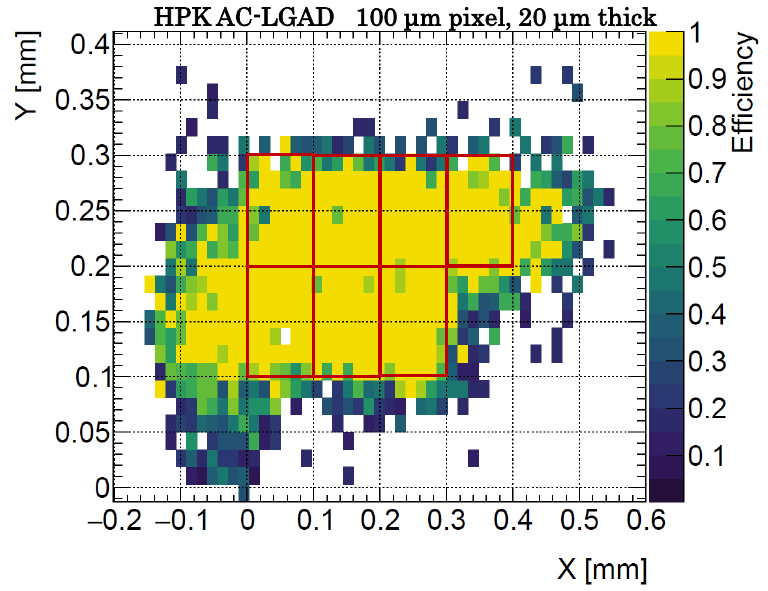}
  \caption{Detection-efficiency map for a 100~\um{}-pitch AC-LGAD pixel sensor measured at the KEK AR test beam line.}
  \label{fig:eff_map}
\end{figure}

Figure~\ref{fig:eff_map} shows the detection-efficiency map measured in the 3~GeV electron beam for the 100~\um-pitch AC-LGAD pixel sensor.

An overall detection efficiency of
\begin{equation}
\epsilon = 99.0 \pm 0.3\%
\end{equation}
was achieved over the active area. The efficiency map shows no significant local inefficiency, including in the regions near the pixel boundaries. This is an important result, because conventional segmented LGADs generally suffer from efficiency loss in the inter-pixel region due to the presence of inactive structures required for segmentation. In contrast, the present measurement demonstrates that the AC-LGAD architecture preserves a nearly uniform detection efficiency across the full sensor area, consistent with its full-fill-factor design.

These results confirm that the pixelated AC-LGAD concept can provide high detection efficiency without introducing dead regions, even at a pixel pitch of 100~\um.

\subsection{Spatial Resolution}

\begin{figure}[htbp]
  \centering
  \includegraphics[width=0.8\linewidth]{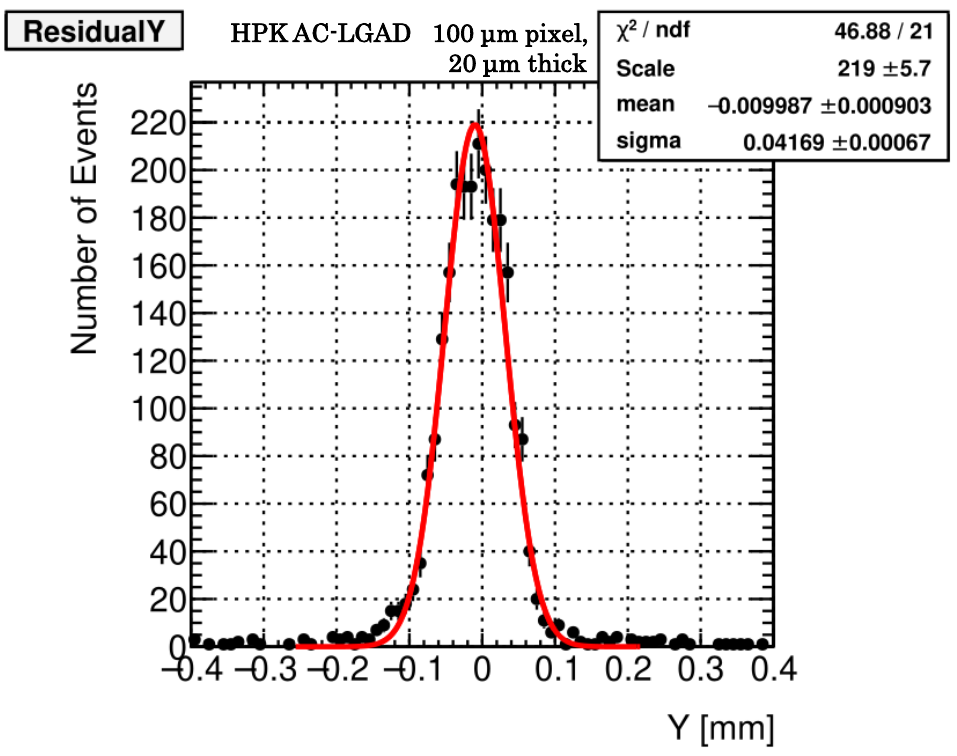}
  \caption{Residual distributions in the $x$ and $y$ directions for the pixelated AC-LGAD detector.}
  \label{fig:residual}
\end{figure}

The spatial resolution was evaluated using the residual distribution between the reconstructed track position and the AC-LGAD hit position.
Figure~\ref{fig:residual} shows the residual distributions in the  $y$ directions.

After subtracting the tracking resolution in quadrature, the intrinsic detector spatial resolution was determined to be
\begin{equation}
\sigma_x = 23.8 \pm 4.7~\um{}, \quad
\sigma_y = 24.9 \pm 4.0~\um{}.
\end{equation}
These values are consistent with the expected resolution for a 100~\um{} pitch detector operated with binary readout.

\subsection{Crosstalk}

\begin{figure}[htbp]
  \centering
  \includegraphics[width=0.9\linewidth]{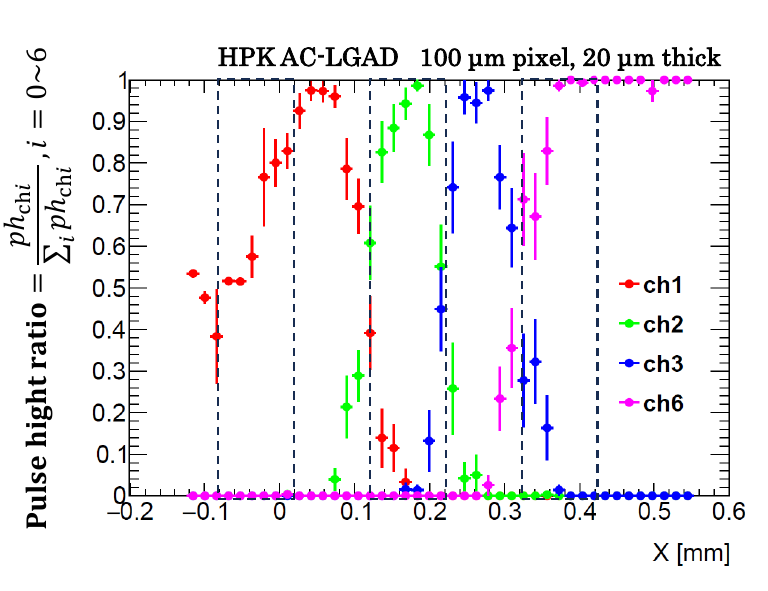}
  \caption{Pulse-height ratio $R_i=ph_i/\sum_{i=0}^{6} ph_i$ as a function of the hit position along the $x$ direction for representative channels of the pixelated AC-LGAD sensor.}
  \label{fig:crosstalk}
\end{figure}

Crosstalk was evaluated using the pulse-height ratio
\begin{equation}
R_i \equiv \frac{ph_i}{\sum_{i=0}^{6} ph_i},
\end{equation}
where $ph_i$ is the pulse height measured in the $i$th channel, and the denominator is the sum of the pulse heights over the seven simultaneously read-out channels. The ratio $R_i$ was studied as a function of the reconstructed hit position along the $x$ direction. Figure~\ref{fig:crosstalk} shows the result for representative channels.

For each channel, the pulse-height ratio becomes nearly unity at the center of the corresponding pixel and falls to nearly zero as the hit position moves into the neighboring pixel. This indicates that, for hits near the pixel center, almost the entire signal is read out by a single pixel channel, while the contribution to adjacent channels remains small. The transition is localized around the pixel boundaries, showing that the crosstalk is well controlled.

These results demonstrate that the pixelated AC-LGAD sensor maintains good channel separation even with AC-coupled readout, which is important for applications in high-occupancy environments.

\section{Conclusion}

We have presented an experimental study of pixelated AC-LGAD detectors with a pixel pitch of 100~\um{} $\times$ 100~\um{}. The results show that the AC-LGAD concept, originally developed to remove the dead regions inherent to conventional segmented LGADs, can be successfully extended to a true pixel geometry.

A timing resolution of $25.3 \pm 0.1$~ps was achieved with a 20~\um{}-thick sensor in the $^{90}$Sr beta-ray measurement, demonstrating that fine pixelation does not significantly degrade the excellent timing capability of LGAD-based sensors. In the 3~GeV electron beam measurement, the observed timing resolution was $40$--$45$~ps, with the difference mainly attributed to the beam-test conditions.

The detector also showed an overall detection efficiency of $99.0 \pm 0.3\%$ without significant efficiency loss at the pixel boundaries, confirming the full-fill-factor nature of the AC-LGAD architecture. The measured intrinsic spatial resolution was about 24~\um{} in both directions, and the crosstalk study showed that the signal is mostly confined to a single pixel except near the pixel boundaries.

These results demonstrate that pixelated AC-LGADs can provide the combination of excellent timing performance, high efficiency, fine spatial resolution, and controlled crosstalk required for future 4D tracking detectors.

\section*{Acknowledgments}

This research was partially supported by Grant-in-Aid for scientific research on advanced basic research (Grant No. 19H05193, 19H04393, 21H0073, 21H01099 and 25H00651) from the Ministry of Education, Culture, Sports, Science and Technology, of Japan as well as the Proposals for the U.S.-Japan Science and Technology Cooperation Program in High Energy Physics from JFY2019 to JFY2027 granted by High Energy Accelerator Research Organization (KEK) and Fermi National Accelerator Laboratory (FNAL). In conducting the present research program, the following facilities have been very important:  electron test beam provided at the AR test beamline (ARTBL) in KEK (Tsukuba).  

\bibliographystyle{elsarticle-num-names} 
\bibliography{LGAD_pub}
\end{document}